\documentclass[conference]{IEEEtran}
\IEEEoverridecommandlockouts

\def\X{{\mathbb X}}

\def \A{{\mathbb A}}
\def \sim{{\textit{sim}}}
\def \R{{\mathbb R}}
\newcommand{\re}[2]{\mathbf r_{{\mathbf #1}_{#2}}}
\newcommand{\ab}[2]{\mathbf e_{{\mathbf #1}_{#2}}}
\newcommand{\se}[2]{\mathbf #1_{#2}}
\usepackage{amsmath}
\usepackage{subcaption}
\DeclareMathOperator*{\argmin}{arg\,min}

\usepackage{cite}
\usepackage{amsmath,amssymb,amsfonts}
\usepackage{graphicx}
\usepackage{textcomp}
\usepackage{mathtools}
\usepackage{xcolor}
\usepackage{balance}
\usepackage{algorithm}
\usepackage{algpseudocode}
\usepackage{caption}
\usepackage{enumitem}

\usepackage{algpseudocode}
\usepackage{multicol}

\begin{document}

\title{Dynamic Relative Representations for Goal-Oriented Semantic Communications \vspace{-.2cm}}

\author{Simone Fiorellino$^{1,2}$, Claudio Battiloro$^{3}$, Emilio Calvanese Strinati$^4$, Paolo Di Lorenzo$^{1,2}$ \smallskip\\ $^1$DIET department, Sapienza University, Rome, Italy. $^2$CNIT, Parma, Italy.\\ $^3$Harvard T.H. Chan School of Public Health, Harvard University, Boston, U.S.A.\\
$^4$CEA Leti, University Grenoble Alpes, 38000, Grenoble, France\\
\normalsize{Emails: \{simone.fiorellino,paolo.dil   orenzo\}@uniroma1.it, cbattiloro@hsph.harvard.edu, emilio.calvanese-strinati@cea.fr} \vspace{-.3cm}
\thanks{This work was supported by the European Union under the Italian National Recovery and Resilience Plan (NRRP) of NextGenerationEU, partnership on “Telecommunications of the Future” (PE00000001 - program “RESTART”), and by the SNS JU project 6G-GOALS under
the EU’s Horizon program Grant Agreement No 101139232.}
}
\maketitle

\begin{abstract}
In future 6G wireless networks, semantic and effectiveness aspects of communications will play a fundamental role, incorporating meaning and relevance into transmissions. However, obstacles arise when devices employ diverse languages, logic, or internal representations, leading to semantic mismatches that might jeopardize understanding. In latent space communication, this challenge manifests as misalignment within high-dimensional representations where deep neural networks encode data. This paper presents a novel framework for goal-oriented semantic communication, leveraging relative representations to mitigate semantic mismatches via latent space alignment. We propose a dynamic optimization strategy that adapts relative representations, communication parameters, and computation resources for energy-efficient, low-latency, goal-oriented semantic communications. Numerical results demonstrate our methodology's effectiveness in mitigating mismatches among devices, while optimizing energy consumption, delay, and effectiveness.
\end{abstract}

\vspace{.1cm}

\begin{figure*}[t!]
    \centering
    \includegraphics[width=0.98\textwidth, trim=0bp 39bp 10bp 0bp, clip]{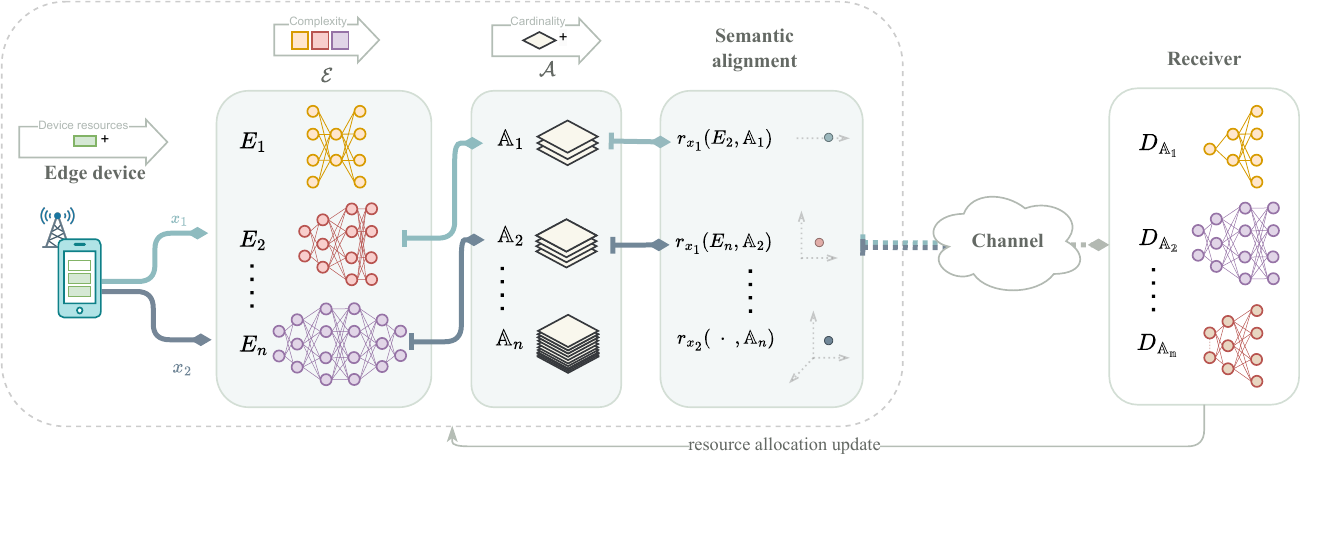}
    \vspace{-0.1cm} 
    \caption{The edge device (TX) encodes data according to the optimal encoder-anchor set pair $(E^*,\mathbb{A}^*)$. The RX is equipped with relative decoders trained on shared anchor sets to interpret the transmitted relative representations.} 
    \label{fig:graphical_abstract}
\end{figure*}

\section{Introduction}

Classic communication paradigms prioritize the accurate transmission of individual symbols or bits without explicit consideration of their underlying meaning. As Weaver identified, communication can be understood through three levels: the syntactic level (focusing on the reliable transmission of signals), the semantic level (concerned with the meaning conveyed by those signals), and the effectiveness level (ensuring the message achieves its intended purpose) \cite{shannon1948mathematical}. 
However, the exponential growth of data-intensive wireless applications demands a fundamental shift in this approach. The sheer volume of information required by applications like autonomous vehicles and augmented reality overwhelms the capacity of traditional bit-centric communication systems.
In contrast, modern networks increasingly demand the exchange of information with shared meaning \cite{strinati20216g}. 
Semantic communication (SemCom) addresses this need by embedding meaning directly into transmissions, moving beyond the mere transmission of bits \cite{guler2018semantic, bao2011towards, gunduz2022beyond,strinati2024goal}. 
To achieve this, SemCom relies heavily on artificial intelligence (AI) and deep neural networks (DNNs) to extract and transmit key semantic features from data rather than the raw data itself. Instead of focusing on the perfect reconstruction of raw symbols, SemCom prioritizes transmitting the underlying meaning, offering inherent robustness to noise and imperfect transmission conditions. At the receiver side, a DNN decoder reconstructs and interprets the intended message. The flexibility and resilience of SemCom make it ideal for emerging applications where devices operate in dynamic and resource-constrained environments \cite{di2023goal}. Building on this idea, novel approaches to semantic joint source and channel coding \cite{gunduz2022beyond}, semantic extraction and compression \cite{kountouris2021semantics, stavrou2023role}, goal-oriented system design and optimization \cite{binucci2022adaptive,di2023goal}, semantic reasoning \cite{thomas2023reasoning}, and semantic communications based on generative AI \cite{barbarossa2023semantic} have been proposed in the literature. Despite these advancements, disruptions caused by misinterpretations of semantic information, known as \textit{semantic noise}, pose a significant challenge. Such disturbances often arise from discrepancies in how different devices represent and interpret data, leading to errors in understanding, even when the physical transmission is error-free. This issue is particularly acute in dynamic edge environments, where various devices are involved, and complex system-wide retraining is not feasible. Misalignments in high-dimensional data representations can severely disrupt semantic understanding under these conditions \cite{sana2023semantic}. This underscores the need for mechanisms that mitigate semantic mismatches without necessitating system updates.
\\
To address these challenges, this paper introduces a novel framework for goal-oriented semantic communication at the wireless network edge based on the recent concept of relative representations (RelReps) \cite{moschella2022relative, norelli2024asif}. Our main contribution is two-fold: (i) We introduce relative representations in SemComs to mitigate the impact of semantic mismatches in dynamic scenarios with devices having heterogeneous logics; (ii) We introduce a dynamic optimization strategy based on Lyapunov stochastic optimization \cite{2010Neely}, which dynamically optimizes communication (i.e., rates), computation (i.e., CPU frequency cycles), and learning resources (i.e., anchor sets, encoders) to enable energy-efficient, low-latency, effective semantic communications. The proposed dynamic resource allocation enjoys simple closed-form solutions, and does not require prior knowledge of the statistics of the random variables affecting the system (e.g., wireless channels). Numerical results assess the performance of our methodology in achieving semantic alignment while striking an optimal trade-off between energy consumption, delay, and goal effectiveness.

\section{Relative Representation Motivations}\label{sec:relrep4semcom}

SemCom aims to exchange the meaning conveyed by data rather than just the raw data itself. To achieve this, transmitters (TXs) and receivers (RXs) typically employ DNNs to encode and decode semantic information, transmitting absolute representations (raw DNN encoder outputs). In this scenario, semantic noise can arise from discrepancies in how different encoders semantically represent the same information due to stochastic factors or design choices. These misalignments introduce misinterpretations at the decoder, disrupting the overall communication objective.
\\
To address this, we introduce RelRep into the SemCom framework, enabling zero-shot stitching – the ability to communicate effectively across diverse encoders without the need for retraining. With RelRep, a TX can switch encoders on the fly, and the RX can still interpret the message without needing to retrain itself for each new encoder, as in Figure \ref{fig:graphical_abstract}. 
The rationale of RelRep for SemCom is explained in the sequel.
\\
\textbf{Relative representation.} Given a training dataset $\X = \{\mathbf{x}_i,y_i\}_i$, the encoding function $E(\cdot)$, e.g., a DNN, maps each sample $\se{x}{i} \in \X$ to its \textit{absolute representation} $\ab{x}{i}:= E(\se{x}{i})\in\mathbb{R}^d$. A subset of $\X$ is defined as the anchors set,  $\A = \{\se{a}{1},\ldots, \se{a}{|\A|}\}$. A generic similarity function $\sim: \mathbb{R}^d \times \mathbb{R}^d \rightarrow \mathbb{R}$ is used to compute the similarity between any two absolute representations. The similarity score $r$ between two such representations is given by $r = sim(\ab{x}{i}, \ab{x}{j})$.
Then, the \textit{relative representation} of a sample $\se{x}{i} \in \X$ is defined as:
\begin{equation}\label{eq:relrep}
    \re{x}{i}(E,\A) := (sim(\ab{x}{i}, \ab{a}{1}),\ldots, sim(\ab{x}{i}, \ab{a}{|\A|}))
\end{equation}
We define $D_\A: \R^{|\A|}\to\R^{\mathcal{T}}$ as a relative decoder function. This is specialized in a downstream task, enabling it to interpret the relationships represented in the relative representation and accurately extract meaning.
\\
\begin{figure}[h] 
    \includegraphics[width=0.9\columnwidth, trim=30bp 10bp 60bp 30bp, clip] {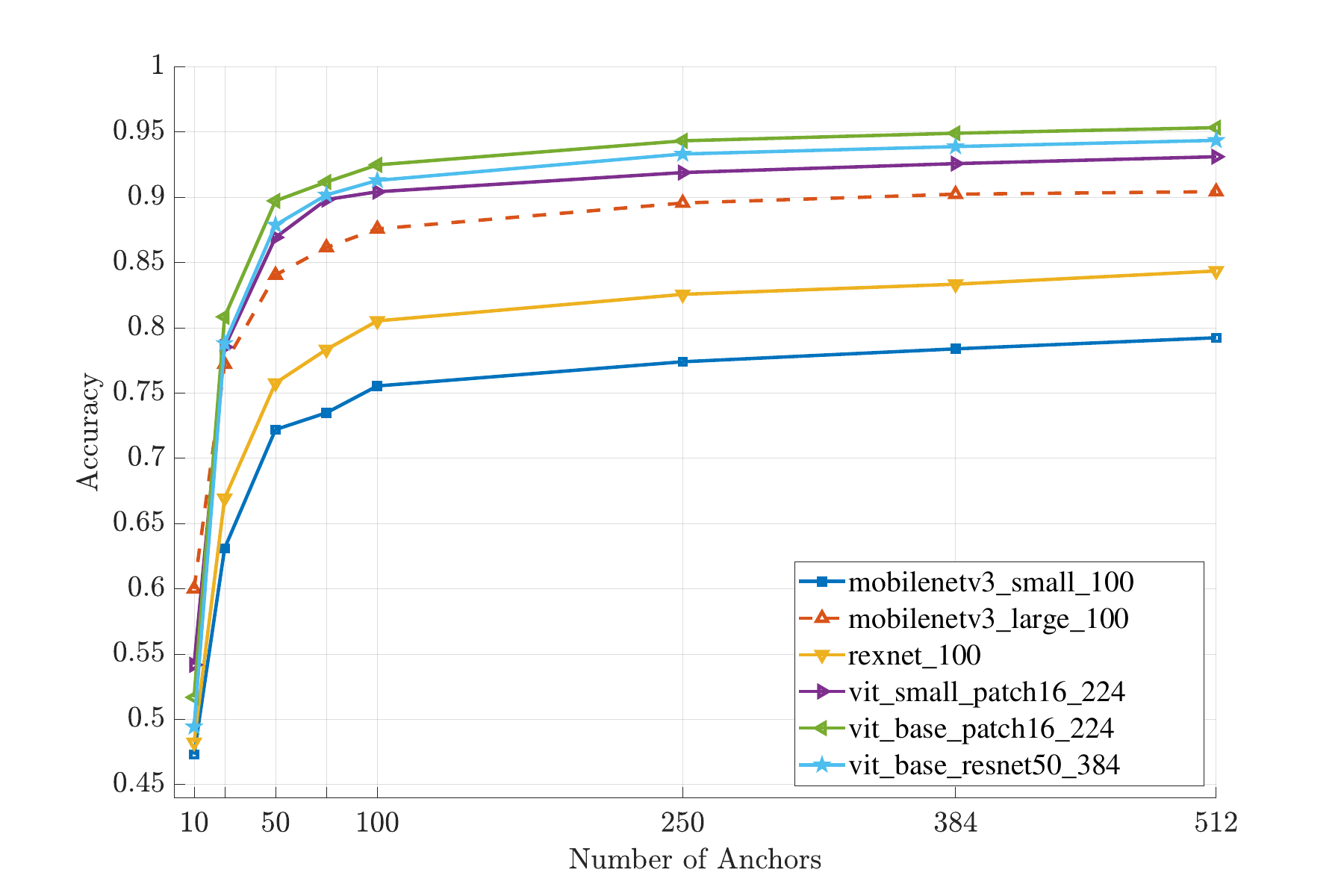} 
    \vspace{-0.1cm} 
    \caption{Performance of zero-shot stitching using different pre-trained encoders and varying anchor set sizes on the CIFAR-10 classification task. The dashed line is end-to-end performance.}
    \vspace{-0.1cm}
    \label{fig:stitched_test_accuracy} 
\end{figure}
\begin{table}[t]
\centering
\begin{tabular}{|lrr|}
\hline
\textbf{Timm model name}  & \textbf{\# Parameters} & \textbf{FLOPs} \\ \hline
mobilenetv3\_small\_100  & 1.5M &  111.98 M \\ \hline
mobilenetv3\_large\_100 & 4.2M & 435.36 M  \\ \hline
rexnet\_100 & 3.5M & 799.28M  \\ \hline
vit\_small\_patch16\_224 & 21.6M & 8.5 G  \\ \hline
vit\_base\_patch16\_224 & 85.7M & 33.72 G  \\ \hline
vit\_base\_resnet50\_384  & 98.1M & 99.04 G \\ \hline
\end{tabular}
\vspace{-0.1cm}
\caption{Timm models used in Fig. \ref{fig:stitched_test_accuracy}.} 
\vspace{-0.7cm}
\label{tab:model_comparison} 
\end{table}
\textbf{Semantic Alignment.} 
RelRep effectively mitigates the effects of semantic mismatches by focusing on data relationships (\ref{eq:relrep}) rather than absolute representations. This approach facilitates semantically equivalent latent spaces and allows dynamic semantic compression based on the number of anchors used. The relative decoders interpret these relationships, enabling zero-shot stitching without retraining, even when encoders are dynamically switched at the edge. This flexibility is demonstrated in Fig. \ref{fig:stitched_test_accuracy}, which shows the performance of model stitching across varying encoding functions and anchor set cardinalities. Notably, zero-shot stitching in absolute cases often fails due to dimensionality mismatches or misalignment, leading to unpredictable behavior across encoders.
In Fig. \ref{fig:stitched_test_accuracy}, we used pre-trained encoders (obtained by removing the final classifier from models trained on ImageNet – see Table \ref{tab:model_comparison} for a full list). The relative decoders were trained using a frozen \textit{mobilenetv3\_large\_100} encoder on the CIFAR-10 dataset. Performance for \textit{mobilenetv3\_large\_100} is therefore end-to-end, while results for other encoders demonstrate zero-shot stitching capabilities. 
In the development of the RelRep design, we followed the approach proposed in \cite{moschella2022relative}. We employed cosine similarity as the similarity function, based on the assumption that the angles between data representations remain consistent across different encoding functions. The anchor sets are generated through a uniform random selection of samples from the training dataset. This methodology ensures a balance between simplicity in implementation and robustness in performance outcomes. Future work may explore optimizing the RelRep design to enhance performance in zero-shot stitching scenarios.
The flexibility enabled by RelRep, through its dynamic semantic compression and encoder agnosticism, is a major advantage in resource-constrained settings, reducing complexity and deployment overhead for edge devices.
Consequently, the semantic understanding workload shifts to the receiver. These advantages motivate the resource optimization strategy we present in the next section, which aims to achieve energy-efficient, goal-oriented semantic communication.

\section{Dynamic Resource Allocation}

In this section, referring to the general communication scheme in Fig. \ref{fig:graphical_abstract}, we devise a dynamic allocation of computation (e.g., CPU clock frequency), communication (e.g., data rate), and learning (e.g., number of anchors, encoders) resources for goal-oriented communication involving inference tasks, while promoting a semantic understanding of the transmitted information. We assume to have sets $\mathcal{E}$ and $\mathcal{A}$ of pre-trained encoders and anchor sets, respectively, available at the TX side.  The RX is assumed to know $\mathcal{A}$, and to have a set of relative decoders (one for each anchor set) to guarantee the semantic alignment between the different pre-trained encoders. The relative decoders are assumed to be trained using the RelReps generated from an arbitrary encoder, which could belong or not to $\mathcal{E}$. In this setting, context parameters like channel quality, data characteristics, and variations in how devices model information (i.e., the semantic language) can fluctuate. Consequently, the proposed dynamic methodology aims to strike an optimal trade-off between power consumption, latency, and inference accuracy by working on the semantic integrity of the communication. The choice of the encoder and the anchor set directly impacts the achievable compression and the computational demands on the TX: Complex models may extract more meaningful information but require higher computational frequencies, consuming more power; similarly, a larger anchor set allows for more nuanced RelReps but increases the volume of data that needs to be transmitted. 

\subsection{System Model}

We assume a simplified RX that incurs negligible decoding delay. Perfect channel state information (CSI) is assumed to be available at the TX. We work with a time-slot-based model, where the system adaptively and dynamically optimizes resource allocation within each time slot. 
In the following, we will describe our system model in detail .\\
\textbf{Power consumption.}  We consider two sources of power consumption for the TX: local computation and transmission. At time $t$, the power spent by the TX for local computation is $p_t^c = \kappa (f_t)^3$,
where $\kappa$ is the effective switched capacitance of the processor and $f_t$ is the CPU clock frequency (Hz) \cite{burd1996processor}. The power spent for  transmission at time $t$ is computed from the Shannon formula as
$$p_t^{u} = \frac{N_0}{|h_t|^2}\Bigg[\exp\Bigg(\frac{R_t}{B}\ln2\Bigg)-1\Bigg],$$
where $B$ is the bandwidth, $|h_t|^2$ the uplink channel power gain, $N_0$ is the noise power spectral density, and $R_t$ is the data rate. Therefore, the total power consumption is given by 
\begin{align}\label{eq:total_power}
    p_t=p_t^{u}+p_t^{c}.
\end{align}
\textbf{Latency.}  The overall latency of each time slot has two main sources of delay: the local processing time and the uplink communication time. The choice of the encoder and the anchor set directly impacts both the local processing time due to the encoder complexity and the uplink communication time due to the size of the relative representation to be transmitted induced by a certain anchor set. The \textit{local processing time} can be computed as $L^c_t = \frac{N_{E_t,\A_t}}{f_t},$ where $N_{E_t,\A_t}$ is the number of CPU cycles necessary to encode the message, and depends on the encoder $E_t \in \mathcal{E}$ and the anchor set $\A_t \in \mathcal{A}$ chosen at time $t$. The \textit{uplink communication time} can be computed as 
$
    L^u_t  = \frac{n_{\A_t} \cdot q}{R_t}
$
where $n_{\A_t} \cdot q$ is the number of bits to be transmitted, and it depends on the size of the relative representation, i.e., the cardinality of the anchor set $\A_t$. Therefore, the total latency is given by
\begin{align}\label{eq:total_latency}
    L_t = L_t^u + L_t^c.
\end{align}
\textbf{Learning Accuracy.} We assume that the RX is provided with a validation set $\X^{\textrm{VAL}}$. A task-dependent function $G_{E,\A}$ is introduced to measure the learning performance, given a pair of encoder and anchor set $E,\A$. As an example, let us consider a classification task, whose validation accuracy can be used to estimate learning performance as
\begin{align}\label{eq:accuracy}
    G_{E,\A} = \frac{1}{|\X^{\textrm{VAL}}|}\sum\nolimits_{x,y  \in \X^{\textrm{VAL}}}\mathbb{I}(\widehat{y}_{E,\A}(\mathbf{x}) = y),
\end{align}
where $\widehat{y}_{E,\A}(\mathbf{x})$ is the prediction for the sample $\mathbf{x}$ if $E \in \mathcal{E}$  and $\A \in \mathcal{A}$ are chosen as encoder and anchor set, respectively. Examples of how the performance metric in (\ref{eq:accuracy}) is affected by different encoders and anchor sets are illustrated in Fig. \ref{fig:stitched_test_accuracy}.

\subsection{Problem Formulation} 
We can now formulate the problem of dynamic resource allocation for goal-oriented communication. The aim is to find the optimal policy to allocate, at each $t$, the uplink data rate $R_t$, the CPU cycles at devices $f_t$, the encoder $E_t$ and the anchor set $\A_t$ to minimize the long-term average system power consumption in (\ref{eq:total_power}), with constraints on the average inference accuracy in \eqref{eq:accuracy} and the average latency in (\ref{eq:total_latency}). 
Then, the dynamic resource allocation problem can be cast as:
\begin{align}\label{dynamic_resource_allocation}
&\min_{R_t, f_t, E_t, \A_t} \quad  \lim_{t\to \infty} \frac{1}{t}\sum_{\tau=0}^{t-1} \mathbb{E} \{p_\tau\} \\
& \textrm{subject to} \quad 
(a) \;  \lim_{t\to \infty} \frac{1}{t}\sum_{\tau=0}^{t-1} \mathbb{E} \{\mathbf{1}_{[L_\tau \geq L^{\textrm{IST}}]}\} \leq p^{\textrm{IST}}; \nonumber \\
& (b) \; \lim_{t\to \infty} \frac{1}{t}\sum_{\tau=0}^{t-1} \mathbb{E} \{L_\tau\} \leq \bar{L};   
\quad (c) \; \lim_{t\to \infty} \frac{1}{t}\sum_{\tau=0}^{t-1} \mathbb{E} \{G_\tau\} \geq \bar{G};  \nonumber  \\
& (d) \;  R^{\textrm{MIN}} \leq R_t \leq R^{\textrm{MAX}}; \quad (e) \;   f_t^{\textrm{MIN}} < f_t \leq f_t^{\textrm{MAX}}; \nonumber \\
& (f) \;  E_t \in \mathcal{E}; \quad (g) \;  \A_t \in \mathcal A. \nonumber
\end{align}
 Expectations are taken over the relevant random variables in the system, i.e., the channel condition and the available computation frequency at the TX side. $G_\tau$ is a short-hand notation for $G_{E_t,\A_t}$ in \eqref{eq:accuracy}. Therefore, we are implicitly assuming that the accuracy evaluated on a validation set is a good proxy for the inference (test) accuracy. This is a standard and reasonable assumption if no shifts in the data distribution happen.
The constraints of the problem (\ref{dynamic_resource_allocation}) have the following meanings: (a) ensures that the probability for the instantaneous latency to exceed a certain value $L^{\textrm{IST}}$ does not exceed a threshold $p^{\textrm{IST}}$; (b) ensures that the mean latency does not exceed a threshold $\bar{L}$; (c) ensures that the mean accuracy of the system does not decrease under a certain threshold $\bar{G}$. Imposing both constraints (a) and (b) is instrumental to better control the statistical distribution of latency. Finally, the constraints in (d)-(g) impose instantaneous bounds (e.g., minimum rates and maximum CPU frequencies) on the resource variables $R_t, f_t$, and impose that $E_t$ and $\A_t$ can take values only from finite sets $\mathcal{E}$ and $\mathcal{A}$ encoders and anchor sets, respectively. 
\vspace{-0.1cm}

\subsection{Algorithmic Solution via Stochastic Optimization} 
We now introduce a dynamic algorithmic framework to solve the long-term optimization problem (\ref{dynamic_resource_allocation}), by recasting it as a stability problem via the tools of stochastic Lyapunov optimization \cite{2010Neely}. First, we define the virtual queues $Z_t$, $Q_t$ and $Y_t$ to enforce the constraints (a)-(b)-(c), respectively, as
\begin{align}
    Y_{t+1} &= \max\{0,Y_t+\epsilon_y(\mathbf{1}_{[L_\tau \geq L^{\textrm{IST}}]} - p^{IST})\} \\
    Z_{t+1} &= \max\{0,Z_t+\epsilon_z(L_t-\bar L)\}\label{eq:latency_vq}\\
    Q_{t+1} &= \max\{0,Q_t+\epsilon_q(\bar G-G_t)\}\label{eq:accuracy_vq}
\end{align}
where $\epsilon_z$, $\epsilon_q$, and $\epsilon_y$ are positive step sizes used to control the convergence speed of the algorithm.
At this point, we define the Lyapunov function as $\mathcal{U}_t = \mathcal{U}(\Phi_t)=\frac{1}{2}(Z_t^2 + Q_t^2 + Y_t^2)$, with $\Phi_t = [Z_t^2, Q_t^2, Y_t^2]$. Then, the drift-plus-penalty function is $\Delta_t^p = \mathbb{E}\{\mathcal U_{t+1}-\mathcal{U}_t + V\cdot p_t^{TOT}|\Phi_t\}$. Now, we want to minimize an upper bound of the drift-plus-penalty \cite{2010Neely}:
\begin{align}
    \Delta_t^p \leq \xi + \mathbb E \Big\{&Z_t(L_t- \bar{L})+Q_t(\bar{G} - G_t)+ \nonumber \\
    &+Y_t(L_t + 1 - p^{IST})+V\cdot p_t^{TOT}|\Phi_t\Big\}
\end{align}
with  $\xi$ a positive finite constant, obtained following a similar procedure to \cite{merluzzi2020dynamic}. 
Finally, using stochastic approximation arguments \cite{2010Neely}, we optimize the upper bound of the drift-plus-penalty, removing the expectation, thus getting the following per-slot problem (omitting all constant terms):
\begin{align}\label{problem:deterministic}
    &\min_{R_t, f_t, E_t, \A_t} \quad  (Z_t + Y_t)L_t - Q_tG_t + V\cdot p_t^{TOT} \nonumber\\
    & \textrm{subject to} \quad
    (f) \;  E_t \in \mathcal{E};\quad 
    (g) \;  \A_t \in \mathcal{A};\nonumber \\
    &(d) \;  R^{MIN} \leq R_t \leq R^{\textrm{MAX}};\quad (e) \; f_t^{\textrm{MIN}} < f_t \leq f_t^{\textrm{MAX}}.
\end{align}
Now, because of (d)-(e)-(f)-(g), \eqref{problem:deterministic} is a mixed-integer nonlinear optimization problem, which might be very complicated to solve. However, for any given $\A_t$ and $E_t$, problem (\ref{problem:deterministic}) becomes a convex problem in $R_t$ and $f_t$. Therefore, we can derive the optimal uplink rate $R_t^*$ and CPU frequency  $f_t^*$ by imposing the KKTs conditions \cite{boyd2004convex}, obtaining
\begin{align}
    R_t^*&=\left[\frac{2B}{\ln2} W\Bigg( \frac{1}{2}\sqrt{\frac{(Z_t+Y_t){n_{\A_t}}q|h_t|^2\ln2}{BVN_0}} \Bigg)\right]_{R^{MIN}}^{R^{MAX}} \label{eq:optimal_rate}\\
    {f_t}^* &= \left[\sqrt[4]{\frac{(Z_t+Y_t)N_t}{3V\kappa}}\right]_{f_t^{MIN}}^{f_t^{MAX}}\label{eq:optimal_fr}
\end{align}
At this point, to determine the optimal encoder and anchor set, the algorithm needs to evaluate the objective function for each possible combination, setting the optimal communication and computation parameters as in (\ref{eq:optimal_rate})-(\ref{eq:optimal_fr}). This means that the system performs $|\mathcal{E}| \cdot |\mathcal{A}|$ independent function evaluations, where $|\mathcal{E}|$ and $|\mathcal{A}|$ are the cardinalities of the encoder and anchor sets, respectively. In this context, the computational cost associated with a single-function evaluation is generally negligible, but greedy approaches could be employed to alleviate the burden further. An overview of the methodology is described in Alg.\ref{alg:resource_optimization}, where we denote as $\Psi(R_t, f_t, E_t, \A_t)$ the objective in problem \eqref{problem:deterministic}. 
\begin{algorithm}[t]
\caption{Resource Optimization for SemCom} 
\label{alg:resource_optimization} 
\textbf{Inputs:} Sample $\mathbf{x}_t$
\\
\textbf{Output:} $E_t^*$, $\A_t^*$, $R_t^*$, $f_t^*$ 
\begin{algorithmic} 
    \For {$t \geq 0$}
        \State (RX) Update the virtual queues $Z_{t}, Q_{t}, Y_{t}$;
        \State (RX) Compute $f_t^*$ and $R_t^*$ as in \eqref{eq:optimal_rate} and \eqref{eq:optimal_fr};
        \State (RX) Set $E_t^*,\A_t^* = \underset{E_t \in \mathcal{E}, \A_t \in \mathcal{A}}{\argmin} \Psi(E_t,\A_t, R_t^*, f_t^*)$;
        \State (TX) Compute and transmit $r_{\mathbf{x}_t}(E^*_t,\A^*_t)$;
    
    \EndFor 
\end{algorithmic}
\end{algorithm}

\begin{figure}[t] 
    \centering
    \includegraphics[width=0.9\columnwidth]{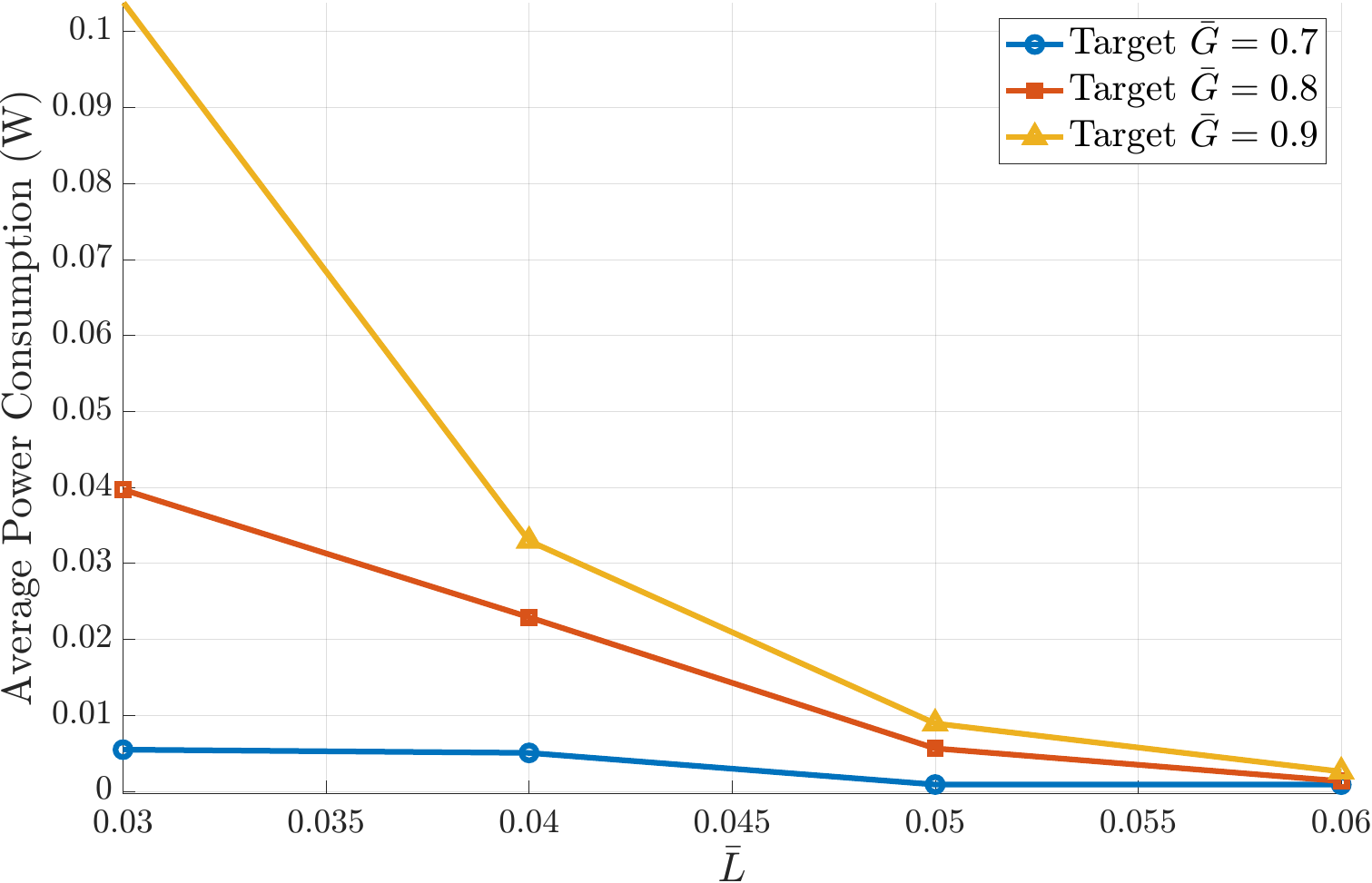} 
    \vspace{-0.15cm} 
    \caption{Long-term average power vs $\bar{L}$, for different $\bar{G}$.} 
    \vspace{-.5cm}
    \label{fig:latency_power_tradeoff}
\end{figure}

\subsection{Numerical Results}

\begin{figure}[h]
\begin{subfigure}{\linewidth}
\centering
   \includegraphics[width=0.95\linewidth]{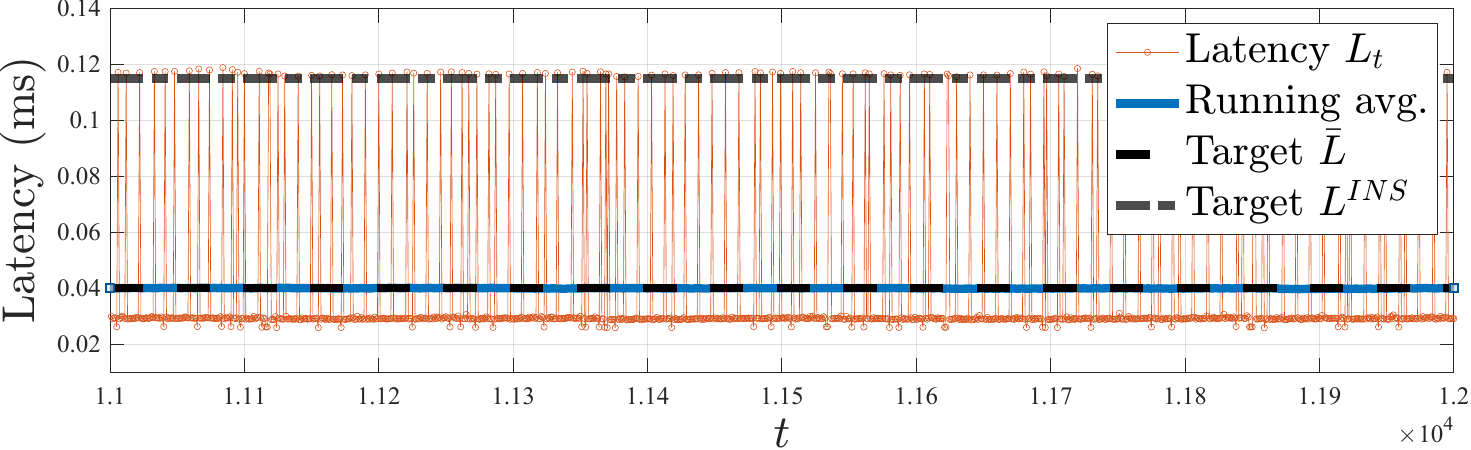}
\end{subfigure}
\begin{subfigure}{\linewidth}
\centering
   \includegraphics[width=0.95\linewidth]{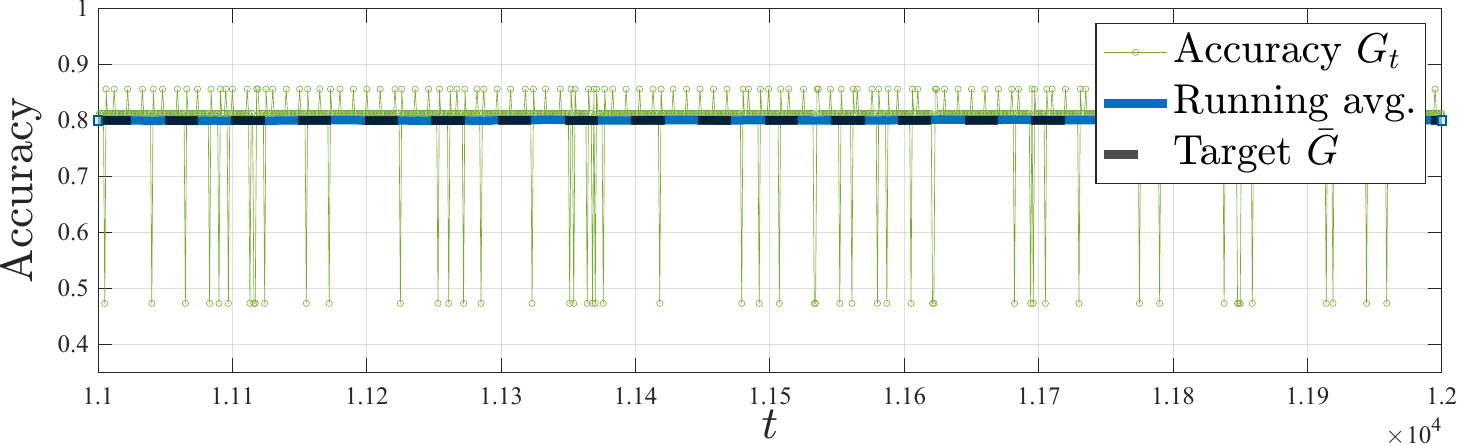}
\end{subfigure}
\vspace{-0.35cm} 
\caption{Latency and accuracy vs $t$, $\bar{L}=0.04\ s$ and $\bar{G}=0.8$.}
\vspace{-.5cm}
\label{fig:inst_behavior}
\end{figure}

To assess the performance of our dynamic resource allocation framework, we run Algorithm \ref{alg:resource_optimization} while varying the constraints on average latency ($\Bar{L}$) and average accuracy ($\Bar{G}$) in (\ref{dynamic_resource_allocation}). This allows us to explore the inherent trade-offs between power consumption, latency, and inference accuracy, highlighting our approach's flexibility. As discussed in Sec. \ref{sec:relrep4semcom}, we focus only on RelReps due to the unpredictable behavior of the absolute case in the zero-shot stitching setting. 
Our simulation setup aligns with the preliminary study in Fig. \ref{fig:stitched_test_accuracy}, employing a set of pre-trained encoders $\mathcal{E}$ (see Table \ref{tab:model_comparison}) and anchor sets $\mathcal{A}$. 
We optimize, across 5 random seeds, according to the following $\bar{L}$ values [0.03,0.04,0.05,0.06] (s) and $\bar{G}$ values [0.7,0.8,0.9].
The channel takes into account path loss propagation, the message quantization ($q$) is set to 32 bits, and the channel bandwidth is $1$ MHz. To ensure timely responses, we define a maximum tolerable latency of $L^{\textrm{IST}}=\Bar{L}+7.5\times 10^{-3}\ s$ with a corresponding probability threshold of $p^{\textrm{IST}}=0.3$. Then, in Fig. \ref{fig:latency_power_tradeoff}, we illustrate the behavior of the average power versus the average latency constraint $\Bar{L}$, for different accuracy constraint $\Bar{G}$. As we can notice from Fig. \ref{fig:latency_power_tradeoff}, RelReps enable the effective exchange of semantic information for all system configurations; as expected, more stringent accuracy constraints necessitate more computationally expensive choices for a given latency constraint. Finally, Fig. \ref{fig:inst_behavior} illustrates the instantaneous behavior of latency (top) and accuracy (bottom) for a specific system configuration, i.e., $\bar{L}=0.04\ s$ and $\bar{G}=0.8$, showing the capability of Algorithm 1 to guarantee the long-term constraints in (\ref{dynamic_resource_allocation}). 

\section{Conclusions}
This paper introduced a novel framework for goal-oriented and SemCom, addressing the challenges posed by semantic mismatches in dynamic environments. Our key contributions lie in using relative representations for robust communication across diverse devices and a dynamic resource optimization strategy for goal-oriented tasks. This strategy adapts communication, computation, and learning resources to enable energy-efficient, low-latency inference crucial for edge applications. Numerical results validate the effectiveness of our framework in achieving a favorable trade-off between energy consumption, delay, and inference reliability.
Future developments include the optimal selection of anchors and similarity functions used in RelReps, with the aim of improving the performance of goal-oriented semantic communications.

\bibliographystyle{IEEEtran}
\bibliography{refs} 

\end{document}